\documentclass[rmp,superscriptaddress,floatfix]{revtex4}

\usepackage{dcolumn}
\usepackage{amsmath}
\usepackage{amssymb}
\usepackage{amsfonts}
\usepackage{graphicx}
\usepackage{bm}   
\usepackage{bbm}   
\usepackage{verbatim}
\usepackage{stmaryrd}
\usepackage{amsthm}

\theoremstyle{break}    \newtheorem*{Def}{Definition}
\theoremstyle{break}    \newtheorem{The}{Theorem}
\theoremstyle{break}    
\theoremstyle{break}    \newtheorem{Cor}{Corollary}

\def\bra#1{\mathinner{\langle{#1}|}}
\def\ket#1{\mathinner{|{#1}\rangle}}

\newcommand{\Lattice}      {\mathrm{L}}

\newcommand{\clrule}      {f}
\newcommand{\Gclrule}      {F}
\newcommand{\States}      {\Sigma}
\newcommand{\Hood}      {\mathcal{N}}

\begin{document}

\title{Quantum Cellular Automata}

\author{Karoline Wiesner}
\email{k.wiesner@bristol.ac.uk}
\affiliation{School of Mathematics, University of Bristol}
\affiliation{Centre for Complexity Sciences, University of Bristol}

\date{\today}

\bibliographystyle{unsrt}

\preprint{arxiv.org/XXX.XXXX}

\maketitle


\tableofcontents


\section{Outline}

Quantum cellular automata (QCA) are a generalization of
(classical) cellular automata (CA) and in particular of reversible CA. The
latter are reviewed shortly. An overview is given over
early attempts by various authors to define one-dimensional QCA. These turned
out to have serious shortcomings which are discussed as well.
 Various proposals subsequently put forward by a number of authors for a general
definition of one- and higher-dimensional QCA are reviewed and their
properties such as universality and reversibility are discussed.

\section{Glossary}

Space-homogeneous : The transition function  / update table is the same for each cell.

Time-homogeneous : The transition function / update table is time-independent.

Configuration : The state of all cells at a given point in time.

Neighborhood : All cells with respect to a given cell that can affect
this cell's state at the next time step. A neighborhood always contains a
finite number of cells.

Update table : Takes the current state of a cell and its neighborhood as an argument and
returns the cell's state at the next time step.

Schr\"odinger picture : Time evolution is represented by a quantum state
evolving in time according to a time-independent unitary operator acting on
it.

Heisenberg picture : Time evolution is represented by 
observables (elements of an operator algebra) evolving in time according
to a unitary operator acting on them.

BQP complexity class :  \emph{Bounded error, quantum probabilistic},
the class of decision problems solvable by a quantum computer in polynomial
time with an error probability of at most 1/3.

QMA complexity class : \emph{Quantum Merlin-Arthur}, the class of
decision problems  such that a "yes" answer can be verified by a 1-message
quantum interactive proof (verifiable in BQP).

Quantum Turing machine : A quantum version of a Turing machine -- an abstract
computational model able to compute any computable sequence. 

Swap operation : The one-qubit unitary gate $ U =  \left(
\begin{array}{cc} 
0 & 1 \\
1 & 0
\end{array} \right)$

Hadamard gate : The one-qubit unitary gate $ U = \frac{1}{\sqrt{2}} \left(
\begin{array}{cc} 
1 & 1 \\
1 & -1
\end{array} \right)$

Phase gate : The one-qubit unitary gate $ U =  \left(
\begin{array}{cc} 
1 & 0 \\
0 & e^{i\phi} 
\end{array} \right)$

Pauli operator : The three Pauli operators are $ \sigma_x =  \left(
\begin{array}{cc} 
0 & 1 \\
1 & 0 
\end{array} \right)$, $ \sigma_y =  \left(
\begin{array}{cc} 
0 & -i \\
i & 0 
\end{array} \right)$, $ \sigma_z =  \left(
\begin{array}{cc} 
1 & 0 \\
0 & -1
\end{array} \right)$

\section{Definition of the subject and its importance}

Quantum cellular automata (QCA) are a quantization of classical cellular
automata (CA),
$d$-dimensional arrays of cells with a finite-dimensional state space and a
local, spatially-homogeneous, discrete-time update rule. For QCA each cell is
a finite-dimensional quantum system and 
the update rule is unitary. CA as well as some versions of QCA have been shown to be
computationally universal. Apart from a theoretical interest in a quantized
version of CA, QCA are a natural framework for what is most
likely going to be the first application of quantum computers -- the
simulation of quantum physical systems.
In particular, QCA are capable of simulating quantum dynamical systems
whose dynamics are 
uncomputable by classical means. 
QCA are now considered one of the standard models of quantum computation next
to quantum circuits and various types of measurement-based quantum
computational models\footnote{For
details on these and other aspects of 
quantum computation see the article by Kendon in this issue.}. Unlike
their classical counterpart,  an axiomatic, all-encompassing definition of
(higher-dimensional) QCA is still missing. 

\section{Introduction}

Automata theory is the study of abstract computing devices and the class of
functions they can perform on their inputs. The original concept of cellular
automata is most strongly associated with John von Neumann ($^* 1903$,
$^\dagger 1957$), a Hungarian
mathematician who made major contributions to a vast range of fields including
quantum mechanics, computer science, functional analysis and many others.
According to Burks, an assistant of von Neumann, \cite{vonN.66} von Neumann had posed the fundamental
questions: ``What kind of logical organization is sufficient for an automaton
to reproduce itself?''. It was Stanislav Ulam who suggested to use the
framework of cellular automata to answer this question. In 1966 von Neumann presented
a detailed analysis of the above question in his book {\it Theory of Self-Reproducing Automata}
\cite{vonN.66}. 

Thus, von Neumann initiated the field of cellular automata. He also made 
central contributions to the mathematical foundations of quantum mechanics
and, in a sense von Neumann's quantum logic ideas were an early attempt at defining
a computational model of physics. But he did not pursue this, and did not go
in the directions that have led to modern ideas of quantum computing  in
general or quantum cellular automata in particular.

The idea of quantum computation is generally attributed to Feynman who, in his now famous
lecture in 1981, proposed a computational scheme based on quantum
mechanical laws \cite{feynman_simulating_1982}. A contemporary
paper by Benioff contains the first proposal of a quantum Turing machine
\cite{benioff_computer_1980}.
 The general idea was to devise a
computational device based on and exploiting quantum phenomena that would
outperform any classical computational device. These first proposals were
sequentially operating  quantum mechanical machines imitating the logical operations of classical
digital computation. The idea of parallelizing the operations was found in
classical cellular automata. However, how to translate cellular automata into a quantum
mechanical framework turned out not to be trivial. And to a certain extent how
to do this in general  remains an open question until today. 

The study of quantum cellular automata (QCA) started with the work of
Gr\"ossing and Zeilinger who coined the term QCA and provided a first
definition
\cite{grossing_quantum_1988}. Watrous developed a different model of QCA
\cite{j._watrous_one-dimensional_1995}. His work lead to further studies by several groups
\cite{dam_quantum, durr_decision_1997, durr_decision_2002}. Independently of this, Margolus
developed a parallelizable quantum computational architecture building on
Feynman's original ideas \cite{margolus_parallel_1991}. For various reasons
to be discussed below, none of these early
proposals turned out to be physical. The study of QCA gained new momentum with the
work by Richter, Schumacher, and Werner
\cite{richter_ergodicity_1996, schumacher_reversible_2004}
and others \cite{perez-delgado_local_2007, arrighi_intrinsically_2007,
arrighi_one-dimensional_2007} who avoided unphysical behavior
allowed by the early proposals \cite{schumacher_reversible_2004,
arrighi_one-dimensional_2007}. It is
important to notice that in spite of the over two-decade long history of QCA
there is no single agreed-upon definition of QCA, in particular of
higher-dimensional QCA. Nevertheless, many useful 
properties have been shown for the various models. Most importantly, quite a
few models are shown to be computationally 
universal, {\it i.e.} they can 
simulate any quantum Turing machine and any quantum circuit efficiently 
\cite{j._watrous_one-dimensional_1995, dam_quantum, raussendorf_quantum_2005,
shepherd_universally_2006, perez-delgado_local_2007}.
Very recently, their ability to generate and transport entanglement has been
illustrated \cite{brennen_entanglement_2003}.

A comment is in order on a class of models which is often labeled as QCA but in
fact are classical cellular automata implemented in quantum mechanical
structures. They do not exploit quantum effects
for the actual computation. To make this distinction clear they are now called
\emph{quantum-dot QCA}. These types of QCA will not be discussed here.

\section{Cellular Automata}
\label{sec.Def}

\begin{Def}[Cellular Automata]
A \emph{cellular automaton} (CA) is a 4-tuple
$(\Lattice,\States,\Hood,\clrule )$ consisting of (1) a $d$-dimensional lattice
of cells $\Lattice$ indexed $i\in \mathbb{Z}^d$, (2) a finite set of states
$\States$, (3) a finite neighborhood scheme $\Hood \subset \mathbb{Z}^d$, and
(4) a local transition function $\clrule : \States^{\Hood} \rightarrow \States$.
\end{Def}

A CA is discrete in time and space. It is \emph{space and time homogeneous} if at each time step 
the same transition function, or \emph{update rule}, is applied simultaneously to all cells.
The update rule is \emph{local} if for a given lattice $\Lattice$ and
lattice site $x$, $\clrule(x)$ is localized in $x + 
\Hood = \{ x+n | x\in \Lattice, n\in \Hood \}$, where $\Hood$ is the
\emph{neighborhood scheme} of the CA. In addition to the locality constraint
the local transition function $\clrule$ must generate a unique global 
transition function mapping a lattice
\emph{configuration} $C_t \in \States^\Lattice$ at time $t$ to a new
configuration $C_{t+1}$ at time $t+1$: $\Gclrule : \States^\Lattice
\rightarrow \States^\Lattice$. 
Most CA are defined on infinite lattices or, alternatively, on finite lattices
with periodic boundary conditions. For finite CA only a finite number of cells
is not in a \emph{quiescent} state, i.e. a state that is not effected by the
update.

The most studied CA are the so-called \emph{elementary CA} -- 1-dimensional
lattices with a set of
two states and a neighborhood scheme of \emph{radius} 1  
(\emph{nearest-neighbor interaction}). i.e. the state of a
cell at point $x$ at time $t+1$ only depends on the state of cells $x-1$, $x$,
and $x+1$ at time $t$.
There are 256
such elementary CA, easily enumerated using a scheme invented by
Wolfram \cite{wolfram_statistical_1983}. As an example and for later
reference, the update table of \emph{rule 110} is given in
Table~\ref{tab.110}. 
 CA with update rule '110' have been shown to be computationally universal,
i.e. they can simulate any Turing machine in polynomial time
\cite{cook_universality_2004}.

\begin{table}[h]
$M^{110} =$ 
\begin{tabular}{cccccccc}
$111$ & $110$ & $101$ & $100$ & $011$ & $010$ & $001$ & $000$ \\
\hline
$0$ & $1$ & $1$ & $0$ & $1$ & $1$ & $1$ & $0$
\end{tabular}
\caption{\label{tab.110} 
Update table for CA rule '110' (the second row is the decimal number '110' in binary notation).
}
\end{table}

A possible approach to constructing a QCA would be to
simply ``quantize'' a CA by rendering the update rule unitary. There are two
problems  with this approach. One is, that applying the same unitary to each
cell does not yield a well-defined global transition function nor necessarily
a unitary one. The second problem is the synchronous update of all cells. 
``In practice'',  the synchronous update of, say, an elementary CA can be achieved by 
storing the current configuration in a temporary register, then update all cells with odd index in
the original CA, update all cells with even index in the register and finally
splice the updated cells together to obtain the original CA at the next time step. 
Quantum states, however, cannot be copied in general due to the so-called
\emph{no-cloning theorem} \cite{wootters_single_1982}. Thus,
parallel update of a QCA in this way is not 
possible. Sequential update on the other hand leads to either an infinite
number of time steps for each update or inconsistencies at the
boundaries. 
One solution is a  partitioning scheme as it is used in the construction of 
reversible CA.

\subsection{Reversible Cellular Automata}
\label{sec.rca}

\begin{Def}[Reversible CA]
A CA is said to be \emph{reversible} if for every current configuration there is
exactly one previous configuration. 
\end{Def}

The global transition function $\Gclrule$ of a reversible CA is \emph{bijective}.
In general, CA are not reversible. Only 16 out of the 256
elementary CA  rules are reversible.
However, one can construct  a reversible CA using a partitioning
scheme developed by Toffoli and Margolus for 2-dimensional CA 
\cite{toffoli_invertible_1990}.

\begin{figure}  
\begin{center}
\resizebox{!}{2.50in}{\includegraphics{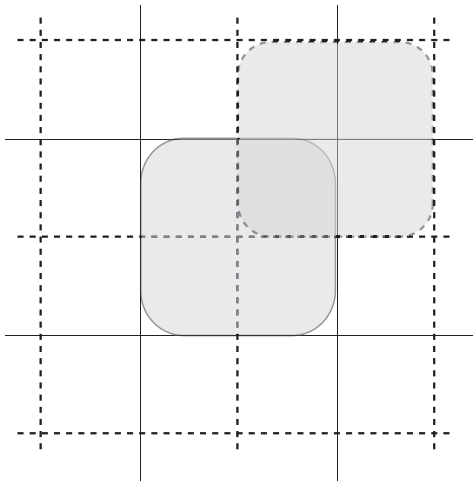}}
\end{center}
\caption{Even (solid lines) and odd (dashed lines) of a Margolus partitioning
scheme in $d=2$ dimensions using blocks of size 
$2\times 2$. For each partition one block is shown shaded. Update rules are
applied alternatingly to the solid and dashed partition.
  }
\label{fig.partition}
\end{figure}

Consider a $2$-dimensional CA with nearest neighborhood scheme $\Hood = \{ x
\in \mathbb{Z}^2 | \forall |x_i| \leq 1\}$.
In the \emph{partitioning scheme} introduced by Toffoli and Margolus each
block of $2\times 2$ cells forms a unit cube $\square$ such that 
the even translates $\square+2x$ with $x\in
\mathbb{Z}^2$ and the odd translates $\square + {\bf 1 } + 2x$,
respectively, form a \emph{partition} of the lattice, see Fig.~\ref{fig.partition}. 
The update rule of a partitioned CA takes as input an entire block of cells
and outputs the updated state of the entire block. The rule is then applied
alternatingly to the even and to the odd translates.
The Margolus partitioning scheme is easily extended to $d$-dimensional
lattices.
A \emph{generalized Margolus scheme} was introduced by Schumacher and Werner
\cite{schumacher_reversible_2004}. It allows for different cell sizes in the
intermediate step. 

A \emph{partitioned CA} is then a CA with a partitioning scheme such that the set
of cells are partitioned in some periodic way: Every cell belongs to exactly
one block, and any two blocks are connected by a lattice translation. 
Such a CA is neither time homogeneous nor space
homogeneous anymore, but periodic in time and space. 
As long as the rule for evolving each block is reversible, the entire
automaton will be reversible.

\section{Early proposals}

Gr\"ossing and Zeilinger were the first to coin the term and
formalize a QCA \cite{grossing_quantum_1988}. 
In the Schr\"odinger picture of quantum mechanics the state of a system at
some time $t$ is described by a state vector $\ket{\psi_t}$ in Hilbert space
$\mathcal{H}$. The state vector evolves unitarily, 
\begin{align}
\ket{\psi_{t+1}} = U \ket{\psi_t}~.
\end{align}
$U$ is a unitary operator, i.e. $UU^\dagger = \bf{1}$, with the complex
conjugate $U^\dagger$ and the identity matrix $\bf{1}$. If $\{\ket{\phi_i}\}$ is a computational basis of the
Hilbert space $\mathcal{H}$ any state $\ket{\psi} \in \mathcal{H}$ can be written as a superposition
$\sum_{\ket{\phi_i}} c_i \ket{\phi_i}$, with coefficients $c_i \in \mathbb{C}$
and $\sum_i c_ic_i^* = 1$. The QCA constructed by Gr\"ossing and Zeilinger is
an infinite $1$-dimensional lattice where at time $t$ lattice
site $i$ is assigned the complex amplitude $c_i$ of state $\ket{\psi_t}$. The
update rule is given by unitary operator $U$. 

\begin{Def}[Gr\"ossing-Zeilinger QCA]
A \emph{Gr\"ossing-Zeilinger QCA} is a 3-tuple $(\Lattice, \mathcal{H}, U)$ which
consists of (1) an infinite 1-dimensional lattice $\Lattice \subseteq
\mathbb{Z}$ representing basis states of (2) a Hilbert space ${\mathcal H}$
with basis set $\{\ket{\phi_i}\}$,
and  (3) a band-diagonal unitary operator $U$.
\end{Def}

Band-diagonality of $U$ corresponds to a locality condition. It turns out that there is no
Gr\"ossing-Zeilinger QCA with nearest-neighbor interaction and  nontrivial
dynamics.
In fact, later on, Meyer showed more generally that ``{\it in one dimension there exists no nontrivial
homogeneous, local, scalar QCA. More explicitly, every band r-diagonal
unitary matrix U which commutes with the one-step translation matrix $T$ is
also a translation matrix $T^k$ for some $k\in \mathbb{Z}$, times a phase}''
\cite{meyer_quantum_1996}. 

Gr\"ossing and Zeilinger also introduce QCA where the unitarity constraint is
relaxed to only
approximate unitarity. After each update the configuration can be  normalized 
which effectively causes non-local interactions.

The properties of Gr\"ossing-Zeilinger QCA are studied by Gr\"ossing and
co-workers in some more 
detail in following years, see \cite{fussy_nonlocal_1993} and references
therein. This pioneering definition of QCA, however, has not been studied much
further, mostly because the ``non-local'' behavior renders the Gr\"ossing-Zeilinger
definition non-physical. In addition, it has little in
common with the concepts developed in quantum computation later on. The
Gr\"ossing-Zeilinger definition really concerns what one would call today
a quantum random walk (for further details see the review by Kempe
\cite{kempe_quantum_2003}). \\

The first model of QCA researched in depth was that introduced by Watrous 
\cite{j._watrous_one-dimensional_1995}, whose ideas where further explored by
van Dam \cite{dam_quantum}, D\"urr, L\^eThanh, and Santha
\cite{durr_decision_1997, durr_decision_2002}, and Arrighi
\cite{arrighi_algebraic_2006}. A Watrous-QCA is defined over an infinite 1-dimensional
lattice, a finite set of states including a quiescent state. The transition
function maps a neighborhood of cells to a single quantum state
instantaneously and simultaneously.
\begin{Def}[Watrous-QCA]
A \emph{Watrous-QCA} is a 4-tuple $(\Lattice, \States, \Hood, f)$ which
consists of (1) a 1-dimensional lattice $\Lattice \subseteq \mathbb{Z}$, (2) a
finite set of cell states $\States$ including a quiescent state $\epsilon$, (3) a finite neighborhood scheme $\Hood$,
and (4) a local transition function $f: \States^\Hood \rightarrow
\mathcal{H}_\States$.
\end{Def}

Here, $\mathcal{H}_\States$ denotes the Hilbert space spanned by the cell states
$\States$. This model can be viewed as a direct quantization of a CA where the
set of possible configurations of the CA is extended to include all linear
superpositions of the classical cell configurations, and the local transition
function now maps the cell configurations of a given neighborhood to a quantum
state. 
One cell is labeled ``accept'' cell.
The quiescent state is used to allow only a finite number of states to be
active and renders the lattice effectively finite. This is crucial to avoid an
infinite product of unitaries and, thus, to obtain a well-defined QCA.

The Watrous QCA, however, allows for non-physical dynamics. It is possible to
define transition functions that do not represent unitary evolution of the
configuration, either by producing superpositions of configurations which do
not preserve the norm, or by inducing a global transition function which is
not unitary. 
This leads to non-physical properties such as super-luminal
signaling \cite{schumacher_reversible_2004}. The set of Watrous QCA is not
closed under composition and inverse 
\cite{schumacher_reversible_2004}.

Watrous defines a restricted class of QCA by introducing a partitioning
scheme.
\begin{Def}[Partitioned Watrous QCA]
A \emph{partitioned Watrous QCA} is a Watrous QCA with $\States = \States_l
\times \States_c \times \States_r$ for finite sets $\States_l$, $\States_c$,
and $\States_r$, and matrix $\Lambda$ of size  $\States \times \States$. For
any state $s = (s_l, s_c, s_r) \in \States$ define 
transition function $f$ as 
\begin{align}
f(s_1, s_2, s_3, s) =  \Lambda_{ (s_{l_3}, s_{m_2}, s_{r_1}), s }~,
\end{align}
with matrix element $\Lambda_{s_i,s_j}$.
\end{Def}

\begin{figure}
\resizebox{!}{1.25in}{\includegraphics{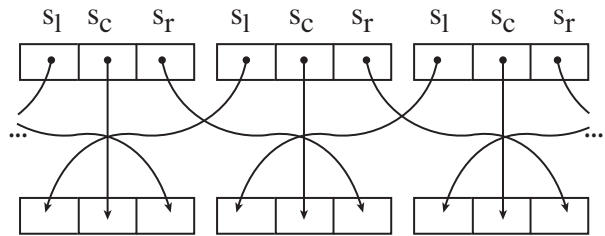}}
\caption{Each cell is divided into three sub-cells labeled l, c, and r for left, center,
and a right, respectively. The
update rule consists of swapping left and right sub-cells of neighboring cells
and then updating each cell internally using a unitary operation acting on the
left, center, and right part of each cell. 
}
\label{fig.watr-qca}
\end{figure}

 In a partitioned Watrous QCA 
each cell is divided into three sub-cells -- left, center, and right. The neighborhood scheme is then a
nearest-neighbor interaction confined to each cell. The
transition function consists of a unitary acting on each partitioned cell
and swap operations among sub-cells of different cells.  
Fig.~\ref{fig.watr-qca} illustrates the swap operation between neighboring
cells. 

For the class of partitioned Watrous QCA Watrous provides the first proof of
computational universality of a QCA
by showing that any quantum Turing machine can be efficiently simulated by
a partitioned Watrous-QCA with constant slowdown and that any partitioned Watrous-QCA
can be simulated by a quantum Turing machine with linear slowdown.

\begin{The}[\cite{j._watrous_one-dimensional_1995}]
Given any quantum Turing machine $M_{TM}$, there exists a partitioned Watrous QCA
$M_{CA}$ which simulates $M_{TM}$ with constant slowdown.
\end{The}

\begin{The}[\cite{j._watrous_one-dimensional_1995}]
Given any partitioned Watrous QCA
$M_{CA}$, there exists a quantum Turing machine $M_{TM}$ which simulates
$M_{CA}$ with linear slowdown. 
\end{The}

Watrous' model was further developed by van Dam \cite{dam_quantum}, who
defines a QCA as an assignment of a product vector to every basis state in the
computational basis. Here the quiescent state is eliminated and thus the QCA
is made explicitly finite. Van Dam showed that
the finite version is also computationally universal.
Efficient algorithms to decide whether a given 1-dimensional QCA is unitary
was presented by D\"urr, LeTanh, and Santha \cite{durr_decision_1997,
durr_decision_2002}.
Due to substantial shortcomings such as non-physical behavior, these early
proposals were replaced by a second wave of proposals to be discussed below.

\section{Definition}

There is not a generally accepted QCA model that has
all the attributes of the CA model: unique definition, simple to describe, and
computationally powerful.  
In particular, there is no axiomatic definition, contrary to its classical
counterpart, that yields an immediate way of constructing/enumerating all of the instances of
this model.  Rather, each set of authors defines QCA in their own particular
fashion. In general, a QCA consists of a $d$-dimensional lattice 
 of identical finite-dimensional quantum systems, a finite set of states, 
a finite neighborhood scheme, and a set of
local unitary transition rules. 
The states $s \in \States$ are basis states spanning a finite-dimensional
Hilbert space. At each point in time a cell represents a finite-dimensional
quantum system in a superposition of  
basis states. The unitary operators represent the discrete-time evolution of
strictly finite propagation speed.

\section{Models of QCA}
\label{sec.mod}

\subsection{Reversible QCA}

Schumacher and Werner use the
Heisenberg picture rather than the Schr\"odinger picture in their model 
\cite{schumacher_reversible_2004}. 
Thus, instead of associating a $d$-level
quantum system with each cell they associate an observable algebra with each
cell. Taking a \emph{quasi-local} algebra as the tensor
product of observable algebras over a finite subset of cells,  a QCA is then a homomorphism of the
quasi-local algebra, which commutes with lattice translations and satisfies 
locality on the neighborhood.

The observable-based approach was first used in
Ref.~\cite{richter_ergodicity_1996} with focus on the irreversible case.
However, this definition left questions open such as whether the composition
of two QCA will again form a QCA. 
The following definition does avoid this uncertainty.

Consider an infinite $d$-dimensional lattice $\Lattice \subset \mathbb{Z}^d$
of cells $x\in \mathbb{Z}^d$, where each cell is associated with the
observable algebra $\mathcal{A}_x$ and each of these algebras is an isomorphic
copy of the algebra of complex $d\times d$-matrices. When $\Lambda \subset
\mathbb{Z}^d$ is a finite subset of cells, denote by $\mathcal{A}(\Lambda)$
the algebra of observables belonging to all cells in $\Lambda$, i.e. the
tensor product $\otimes_{x\in \Lambda} \mathcal{A}_x$. The completion of this
algebra is called a \emph{quasi-local} algebra and will be denoted by
$\mathcal{A}(\mathbb{Z}^d)$.

\begin{Def}[Reversible QCA]
A \emph{quantum cellular automaton} with neighborhood scheme $\Hood \subset
\mathbb{Z}^d$ is a homomorphism $T : \mathcal{A}(\mathbb{Z}^d) \rightarrow
\mathcal{A}(\mathbb{Z}^d)$ of the quasi-local algebra, which commutes with
lattice translations, and satisfies the locality condition
$T(\mathcal{A}(\Lambda)) \subset T(\mathcal{A}(\Lambda + \Hood)) $ for every
finite set $\Lambda \subset \mathbb{Z}^d$. The local transition rule of a
cellular automaton is the homomorphism $T_0 : \mathcal{A}_0 \rightarrow
\mathcal{A}(\Hood)$.
\end{Def}

\begin{figure}
\resizebox{!}{1.25in}{\includegraphics{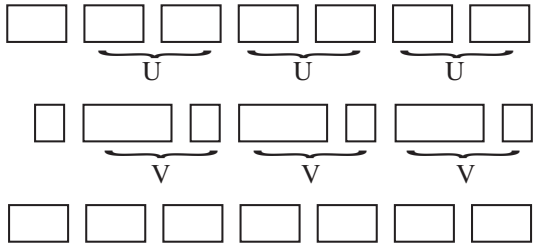}}
\caption{Generalized Margolus partitioning scheme in $1$ dimension using two
unitary operations $U$ and $V$. 
}
\label{fig.revers-qca}
\end{figure}

They present and prove the following theorem on one-dimensional QCA. 

\begin{The}[Structure Theorem \cite{schumacher_reversible_2004}]
\label{the.str}
Let $T$ be the global transition homomorphism of a one-dimensional nearest-neighbor QCA
on the lattice $\mathbb{Z}^d$ with single-cell algebra $\mathcal{A}_0 =
\mathcal{M}_d$. Then $T$ can be represented in the generalized Margolus
partitioning scheme, i.e. $T$ restricts to an isomorphism 
\begin{align}
T : \mathcal{A}(\square) \rightarrow \bigotimes_{s \in \States} \mathcal{B}_s~,
\end{align}
where for each quadrant vector $q \in Q$, the subalgebra  $\mathcal{B}_q
\subset \mathcal{A}(\square + q)$ is a full matrix algebra, $\mathcal{B}_q
\mathcal{M}_{n(q)}$. These algrebras and the matrix dimensions n(q) 
are uniquely determined by $T$.
\end{The}

Theorem~\ref{the.str} does not hold in higher dimensions \cite{werner_private}.

A central result obtained in this framework is that almost any
\cite{werner_private} $1$-dimensional QCA can be
represented using a set of local unitary operators and a generalized
Margolus partitioning \cite{schumacher_reversible_2004}, as illustrated in
Fig.~\ref{fig.revers-qca}. 
Furthermore, if the local implementation allows local ancillas, 
 then any QCA, in any lattice dimension can be built from local unitaries
\cite{schumacher_reversible_2004,werner_private}.
In addition, they prove the following Corollary.
\begin{Cor}[\cite{schumacher_reversible_2004}]
The inverse of a nearest-neighbor QCA exists, and is a nearest-neighbor QCA.
\end{Cor}

The latter result is not true for CA. A similar result for finite
configurations is obtained in \cite{arrighi_one-dimensional_2007}. Here
evidence is presented that the result does not hold for two dimensional QCA.
The work by Schumacher and Werner can be considered the first general
definition for $1$-dimensional QCA. A similar result for many-dimensional QCA
does not exist.

\subsection{Local unitary QCA}

P\'eres-Delgado and Cheung propose a \emph{local unitary
QCA} \cite{perez-delgado_local_2007}. 

\begin{Def}[Local-unitary QCA]
A \emph{local-unitary QCA} is a 5-tuple $\{ (\Lattice,\States,\Hood,U_0,V_0) \}$
consisting of (1) a $d$-dimensional lattice of cells indexed by integer tuples $\Lattice
\subset \mathbb{Z}^d$, (2) a finite set of orthogonal basis states $\States$, (3) a
finite neighborhood scheme $\Hood \subseteq \mathbb{Z}^d$, (4) a local read
function $U_0: (\mathcal{H}_\States)^{\otimes \Hood} \rightarrow
(\mathcal{H}_\States)^{\otimes \Hood}$, and (5) a local update function $V_0 :
\mathcal{H}_\States \rightarrow \mathcal{H}_\States$. The read operation
carries the further 
restriction that any two lattice translations $U_x$ and $U_y$ must commute for
all $x,y \in \Lattice$.
\end{Def}

The product $VU$ is a valid local, unitary
quantum operation. The resulting global update rule is well defined and space
homogeneous. The set of states includes a {\it quiescent} state
as well as an ``ancillary'' set of states  / subspace which can store the result of the
``read'' operation. The initial state of a local-unitary QCA consists of
identical $k^d$ blocks of cells initialized in the same state.
Local-unitary QCA are universal in the sense that for any
arbitrary quantum circuit there is a local-unitary QCA which can simulate it. In addition 
any local-unitary QCA can be simulated efficiently using a family of quantum
circuits \cite{perez-delgado_local_2007}.
Adding an additional memory register to each cell allows this class of QCA to
model any reversible QCA of the Schumacher/Werner type discussed above.

\subsection{Block-partitioned and nonunitary QCA}

Brennen and Williams introduce a model of QCA which allows for unitary and
nonunitary rules 
\cite{brennen_entanglement_2003}. 

\begin{Def}[Block-partitioned QCA]
A \emph{Block-partitioned QCA}  is a 4-tuple $\{\Lattice, \States, \Hood, M\}$
consisting of (1) a $1$-dimensional 
lattice of $n$ cells indexed $\Lattice = 0, \dots, n-1$, (2) a $2$-dimensional
state space $\States$, (3) a neighborhood scheme $\Hood$, and (4) an update
rule $M$ applied over $\Hood$.
\end{Def}

Given a system with nearest-neighbor interactions, the simplest unitary QCA
rule has radius $r=1$ describing a unitary operator applied over a three-cell
neighborhood $j-1,j,j+1$:
\begin{align}
M(u_{00},u_{01},u_{10},u_{11}) = \ket{00}\bra{00}\otimes u_{00} +
\ket{01}\bra{01} \otimes u_{01} + \ket{10}\bra{10}\otimes u_{10} +
\ket{11}\bra{11}\otimes u_{11}~,
\end{align}

where $\ket{ab}\bra{ab}\otimes u_{ab}$ means update the qubit at site $j$ with
the unitary $u_{ab}$ if the qubit at the site $j-1$ is in state $\ket{a}$ and
the qubit at site $j+1$ is in state $\ket{b}$. $M$ commutes with its own
2-site translation. Thus, a partitioning is
introduced by updating simultaneously all even qubits with rule $M$ before
updating all odd qubits with rule $M$. 
Periodic boundaries are assumed. However, by addressability
of the end qubits simulation of a block-partitioned QCA by a QCA with
boundaries can be achieved.

Nonunitary update rules
correspond to completely positive maps on the quantum states where the
neighboring states act as the environment. Take a nearest-neighbor
$1$-dimensional Block-partitioned QCA. In the density operator formalism
each quantum system $\rho$ is given by the probability distribution $\rho
= \sum_i p_i \ket{\psi}\bra{\psi}$ over outer products of quantum states
$\ket{\psi}$. A completely positive map $S(\rho)$ applied to state $\rho$ is
represented by a set of Krauss operators $F_\mu$, which are positive
operators that sum up to the identity $\sum_\mu F_\mu^{\dagger}F_\mu = {\bf 1}$. The
map $S_j^{ab}(\rho)$ acting on cell $j$ conditioned on state $a$ of the left neighbor and state $b$
of the right neighbor can then be written as
\begin{align}
S_j^{ab}(\rho) = \ket{ab}\bra{ab} \otimes \sum_\mu F_\mu^{ab} \rho
F_\mu^{ab\dagger} \otimes \ket{ab}\bra{ab}~.
\end{align}

As an example, the CA rule '110' can now be translated into an update rule for
cell $j$ in a block-partitioned nonunitary QCA:
\begin{align}
F_1^j &= \ket{00}\bra{00}\otimes {\bf 1}^j + \ket{10}\bra{10}\otimes {\bf 1}^j
+ \ket{11}\bra{11}\otimes \sigma_x^j + \ket{01}\bra{01}\otimes
\ket{1}_{jj}\bra{1} \\
F_2^j &= \ket{01}\bra{01}\otimes \ket{1}_{jj}\bra{0}~,
\end{align}

where $\sigma_x$ is the Pauli operator.

The implementation of such a block-partitioned nonunitary QCA is proposed in form of a lattice of even
order constructed with an alternating array of two distinguishable species
$ABABABAB\dots$ that are globally addressable and interact via the Ising
interaction.
Update rules that generate and distribute entanglement are studied in this
framework \cite{brennen_entanglement_2003}.

\subsection{Continuous-time QCA}

Vollbrecht and Cirac initiate the study of continuous-time QCA
\cite{vollbrecht_quantum_2008}. They show that the computability of the ground state
energy of a translationally invariant $n$-neighbor Hamiltonian is QMA-hard.
Their QCA model is taken up by Nagaj and Wocjam 
\cite{nagaj_hamiltonian_2008} who use the term \emph{Hamiltonian QCA}. 

\begin{Def}[Hamiltonian QCA]
A \emph{Hamiltonian QCA} is a tuple $\{ (\Lattice,\States = \States_p \times
\States_d,) \}$
consisting of (1) a $1$-dimensional lattice of length $L$, 
 (2) a finite set of orthogonal basis states $\States =
\States_p \times \States_d$ containing (2a) a data
register $\States_d$, and (2b) a program register $\States_p$.
\end{Def}

The initial state encodes both the program
and the data, stored in separate subspaces of the state space:
\begin{align}
\ket{\phi} = \bigotimes_{j=1}^{L} \left( \ket{p_j} \otimes \ket{d_j} \right)_j
\end{align}

 The computation is
carried out autonomously. Nagaj and Wocjam show that, if the 
system is left alone for a period of time 
$t = O(L~ log\, L)$, polynomially in the length of the chain, 
the result of the computation is obtained with probability $p \geq 5/6
- O(1/log \,L)$. Hamiltonian QCA are computationally universal, more
precisely they are in the complexity class BQP. 
Two constructions for Hamiltonian QCA are given in
\cite{nagaj_hamiltonian_2008}, one using a $10$-dimensional state space, and
the resulting system can be thought of as the diffusion of a system of free
fermions. The second construction given uses a $20$-dimensional state space
and can be thought of as a quantum walk on a line.

\subsection{Examples of QCA}

Raussendorf presents an explicit construction of QCA and proves its
computational universality \cite{raussendorf_quantum_2005}. The QCA lives on a
torus with a $2\times 2$ Margolus partitioning. The update rule is given by a
single 4-qubit unitary acting on $2\times 2$ blocks of qubits. The four-qubit unitary operation
consists of swap operations, the Hadamard transformation, and  a phase
gate.
 The initial state of the QCA is prepared such that columns encode
alternatingly data and program. When the QCA is running the data travel
in one direction while the program 
(encoding classical information in orthogonal states) travels
in the opposite direction. Where the two cross the computation is carried out
through nearest-neighbor interaction. After a fixed number of steps the
computation is done and the result can be read out of a dedicated ``data''
column. This QCA is computationally universal, more precisely, it is within a
constant as efficient as a quantum logic network with local and nearest-neighbor
gates.

Shepherd, Franz, and Werner compare \emph{classically controlled} QCA to autonomous QCA
\cite{shepherd_universally_2006}.
The former is controlled by a classical compiler that selects a sequence of
operations acting on the QCA at each time step. The latter operates
autonomously, performing the same unitary operation at each time step. The
only step that is classically controlled is the measurement (and
initialization). They show the computational equivalence of the two models.
Their result implies that a particular quantum simulator may be as powerful as
a general one.

\section{Computationally universal QCA}

Quite a few models have been  shown to be computationally 
universal, {\it i.e.} they can 
simulate any quantum Turing machine and any quantum circuit efficiently.
A  Watrous-QCA simulates any quantum Turing machine with constant slowdown
\cite{j._watrous_one-dimensional_1995}. The QCA defined by Van Dam is a finite
version of a Watrous QCA and is computationally universal as well
\cite{dam_quantum}. Local-unitary QCA can simulate any quantum circuit and
thus are computationally universal \cite{perez-delgado_local_2007}.
Block-partitioned QCA can simulate a quantum computer with linear 
overhead in time and space \cite{brennen_entanglement_2003}. Continuous-time
QCA are in complexity class BQP and thus computationally universal
\cite{vollbrecht_quantum_2008}. The explicit constructions of $2$-dimensional
QCA by Raussendorf is computationally universal, more precisely, it is within a
constant as efficient as a quantum logic network with local and nearest-neighbor
gates \cite{raussendorf_quantum_2005}.
Shepherd, Franz, and Werner provide an explicit construction of a $12$-state
$1$-dimensional QCA which is in complexity class BQP. 
It is universally programmable in the sense that it
simulates any quantum-gate circuit with polynomial overhead 
\cite{shepherd_universally_2006}. 
Arrighi and Fargetton propose a $1$-dimensional QCA capable of simulating any
other $1$-dimensional QCA with linear overhead
\cite{arrighi_intrinsically_2007}. 

Implementations of computationally universal QCA have been suggested by Lloyd
\cite{lloyd_potentially_1993} and Benjamin \cite{benjamin_quantum_2001}.

\section{Modeling Physical Systems}

One of the goals in developing QCA is to create a useful modeling tool for
physical systems. Physical systems that can be simulated with QCA include Ising and
Heisenberg interaction spin chains, solid state NMR, and quantum lattice
gases. 
Spin chains are perhaps the most obvious systems to model with QCA. The simple
cases of such $1$-dimensional lattices of spins are Hamiltonians which commute
with their own lattice translations. Vollbrecht and Cirac have shown that the
computability of the ground state 
energy of a translationally invariant $n$-neighbor Hamiltonian is in
complexity class QMA \cite{vollbrecht_quantum_2008}.
For simulating noncommuting Hamiltonians a block-wise update such as the
Margolus partitioning has to be used (see Section~\ref{sec.rca}). 
Here the fact is used that
any Hamiltonian can be expressed as the sum of two Hamiltonians, $H = H_a +
H_b$. $H_a$ and $H_b$ can then, to a good approximation, be applied
sequentially to yield the original Hamiltonian $H$, even if these do not
commute. 
It has been shown that such $1$-dimensional spin chains can be simulated
efficiently on a classical computer \cite{vidal_efficient_2004}. It is not known,
however, whether higher dimensional spin systems can be simulated efficiently
classically.

\subsection{Quantum lattice gas automata}

Any numerical evolution of a discretized partial differential equation can be interpreted
as the evolution of some CA, using the framework of \emph{lattice gas
automata}. In the continuous time and space limit 
such a CA mimics the behavior of the partial differential equation. In quantum
mechanical lattice gas automata (QLGA) the
continuous limit on a set of so called quantum lattice Boltzman equation
recovers the Schr\"odinger equation \cite{succi_lattice_1993}.
The first formulation of a linear unitary CA is given in Ref.~\cite{bialynicki-birula_weyl_1994}.
Meyer coins the term \emph{quantum lattice gas automata} (QLGA) 
and demonstrates the equivalence of
a QLGA and the evolution of a set of quantum lattice Boltzman equations
\cite{meyer_quantum_1996, meyer_absence_1996}.
Meyer \cite{meyer_quantum_1997}, Boghosian and Taylor
\cite{boghosian_quantum_1998}, and Love
and Boghosian \cite{love_dirac_2005} explored the idea of
using QLGA as a model for simulating physical systems.
Algorithms for implementing QLGA on a quantum computer have been presented in
\cite{boghosian_simulating_1998, ortiz_quantum_2001, meyer_quantum_2002}.

\section{Implementations}
\label{sec.impl}

A large effort is being made in many laboratories around the world to implement a
model of a quantum computer. So far all of them are confined to a very finite
number of elements and are no way near to a quantum Turing machine (which in
itself is a purely theoretical construct but can be approximated by a very
large number of computational elements). One existing experimental set-up that
is very promising for quantum information processing and that does not suffer
from this ``finiteness'' are optical lattices (for a review, see
\cite{bloch_ultracold_2005}). They possess a translation symmetry which makes QCA
a very suitable framework in which to study their computational power. Optical
lattices are artificial crystals of 
light and consist of hundreds of thousands of microtraps. One or more neutral atoms
can be trapped in each of the potential minima. If the potential minima are 
deep enough any tunneling between the traps is suppressed and each site
contains the same amount of atoms. A quantum register -- here in form of a so-called Mott
insulator -- has been created. The biggest challenge at the moment is to find
a way to address the registers individually to implement quantum gates. 
For a QCA all that is needed is implementing the unitary operation(s) acting
on the entire lattice simultaneously. The internal structure of the QCA
guarantees the locality of the operations. This is a huge simplification
compared to individual manipulation of the registers.
Optical lattices are created routinely by superimposing two or
three orthogonal standing waves generated from laser beams of a certain
frequency. They are used to study Fermionic and
Bosonic quantum gases, nonlinear quantum dynamics, strongly correlated quantum
phases, to name a few topics. 

A type of locally addressed architecture by global control was put forward by Lloyd
\cite{lloyd_potentially_1993}. In this scheme a 1-dimensional array is built
out of three atomic species, periodically arranged as $\mathcal{ABCABCABC}$. Each
species encodes a qubit and can be manipulated without affecting the 
other species. The operations on any species can be controlled 
by the states of the neighboring cells. The end-cells are
used for readout, since they are the only individually addressable components.
Lloyd showed that such a quantum architecture is universal.
Benjamin investigated the minimum physical requirements for such a
many-species implementation and  found a similar architecture using only two types of species, again
arranged periodically $\mathcal{ABABAB}$ \cite{benjamin_schemes_2000,
benjamin_quantum_2001, benjamin_quantum_2004}. By
giving explicit sequences of operations implementing one-qubit and two-qubit
(CNOT) operations Benjamin shows computational universality.
But the reduction in spin resources comes with an increase in logical
encoding into four spin sites with a buffer space of at least four empty spin
sites between each logical qubit.

A continuation of this multi-species QCA architecture is found in the work by
Twamley \cite{twamley_quantum-cellular-automata_2003}. Twamley constructs a
proposal for a QCA architecture based on Fullerene ($C_{60}$) molecules doped with
atomic species $^{15}N$ and $^{31}P$, respectively, arranged alternatingly in a one-dimensional
array. Instead of electron spins which would be too
sensitive to stray electric charges the quantum information is
encoded in the nuclear spins. Twamley constructs sequences of 
pulses implementing 
Benjamin's scheme for one- and two-qubit operations. The weakest point of the
proposal is the readout operation which is not well-defined.

A different scheme for implementing QCA was suggested by T\'oth and Lent
\cite{toth_quantum_2001}. Their scheme is based on the technique of quantum-dot CA.
The term quantum-dot CA is usually used for CA implementations in quantum
dots (for classical computation). The authors, therefore, call their model a
\emph{coherent} quantum-dot CA. They illustrate the usage of an array of N
quantum dots as an N-qubit quantum register. However, the set-up and the
allowed operations allow for individual control of each cell. This coherent
quantum-dot CA is more a hybrid of a quantum circuit with individual qubit
control and a QCA with constant nearest-neighbor interaction. The main
property of a QCA, operating under global control only, is not taken advantage
of.

\section{Future directions}

The field of QCA is developing rapidly. New definitions have appeared very
recently. Since QCA are now considered to be one of the standard
measurement-based models of quantum computation, further work on a consistent
and sufficient definition of higher-dimensional QCA is to be expected. One
proposal for such a ``final'' definition has been put forward in Refs.
\cite{arrighi_one-dimensional_2007, arrighi_n-dimensional_2007}.

In the search for a robust and
easily implementable quantum computational architectures QCA are of quite some
interest. 
The main strength of QCA is global control without the need
to address cells individually (with the possible exception of the read-out
operation). It has become clear that the global update of a QCA would be a way around 
practical issues related
to the implementation of quantum registers and the difficulty of their individual
manipulation.

More concretely, QCA provide a 
natural framework for describing quantum dynamical evolution of optical
lattices, a field in which the experimental physics community has made huge progress
in the last decade.

The main focus so far has been on reversible QCA. Irreversible QCA are closely
related to measurement-based computation and remain to be explored further.

\section{Secondary references}

Summaries of the topic of QCA can be found in Chapter 4.3 of Gruska
\cite{gruska_quantum_1999}, and in Refs. \cite{perez-delgado_models_2005,
aoun_introduction_2004}.



\bibliography{quantum-cellular-automata}

\end{document}